\documentclass{TIIS}
\makeatletter
\AtBeginDocument{%
  \@ifundefined{SI}{}{%
    \@ifundefined{RenewCommandCopy}{\let\qty\SI}{\RenewCommandCopy\qty\SI}%
  }%
}%
\makeatother
\usepackage{cite}

\usepackage{amssymb,amsfonts}
\usepackage{algorithmic}
\usepackage{graphicx}
\DeclareRobustCommand{\maybeincludegraphics}[2][]{%
  \IfFileExists{#2}{\includegraphics[#1]{#2}}{%
    \fbox{\scriptsize Missing file: {\ttfamily\detokenize{#2}}}%
  }%
}
\usepackage{textcomp}
\usepackage{url}
\usepackage{bm}
\usepackage{algorithm}
\usepackage{subcaption}
\usepackage{tikz}
\usetikzlibrary{positioning,arrows}
\usepackage{stfloats}
\usepackage{verbatim}
\usepackage{booktabs}
\usepackage{hhline}
\usepackage[dvipsnames]{xcolor}
\usepackage{pifont}
\usepackage{wrapfig}
\usepackage[colorlinks=true,
            linkcolor=TIISPureBlue,
            citecolor=red,
            urlcolor=blue]{hyperref}
\usepackage[normalem]{ulem}
\usepackage{cleveref}
\usepackage{colortbl}
\definecolor{LightGray}{gray}{0.92}
\usepackage{etoolbox}
\usepackage{array,tabularx}
\usepackage{placeins}
\usepackage{float}
\newcolumntype{Y}{>{\centering\arraybackslash}X}

\makeatletter
\pretocmd{\@citex}{\begingroup\color{red}}{}{}
\apptocmd{\@citex}{\endgroup}{}{}
\makeatother

\DeclareRobustCommand{\CrossRefArticle}[1]{%
  \textcolor{blue}{\uline{\href{#1}{Article (CrossRef Link)}}}%
}

\makeatletter
\let\oldthebibliography\thebibliography
\renewcommand{\thebibliography}[1]{%
  \oldthebibliography{#1}%
  \renewcommand{\@biblabel}[1]{\hbox to\labelwidth{[##1]\hfil}}
  \setlength{\itemsep}{0.5pt}%
  \setlength{\parskip}{0pt}%
  \setlength{\baselineskip}{11.5pt}%
}
\makeatother



\newcommand{\figref}[1]{%
  \hyperref[#1]{\textcolor{TIISPureBlue}{Fig.~\ref*{#1}}}%
}

\newcommand{\tabref}[1]{%
  \hyperref[#1]{\textcolor{TIISPureBlue}{Table~\ref*{#1}}}%
}

\newcolumntype{C}[1]{>{\centering\arraybackslash}m{#1}}
\newcolumntype{L}[1]{>{\raggedright\arraybackslash}m{#1}}
\newcolumntype{R}[1]{>{\raggedleft\arraybackslash}m{#1}}

\def\BibTeX{{\rm B\kern-.05em{\sc i\kern-.025em b}\kern-.08em
    T\kern-.1667em\lower.7ex\hbox{E}\kern-.125emX}}

\renewcommand{\tiisvol}{3}
\renewcommand{\tiisno}{6}
\renewcommand{\tiisdate}{December 20XX}
\renewcommand{\tiisrunningauthors}{Bae}
\renewcommand{\tiisrunningtitle}{ASTRA: Mapping Art-Technology Institutions via Conceptual Axes, Text Embeddings, and Unsupervised Clustering}

\renewcommand{\TIISdoi}{http://doi.org/10.3837/tiis.2020.0X.00X}
\renewcommand{\TIISissn}{\textbf{ISSN : 1976-7277}}
\renewcommand{\TIISfirstpagefootnote}{}

\begin{document}

\title{ASTRA: Mapping Art-Technology Institutions via Conceptual Axes, Text Embeddings, and Unsupervised Clustering}
\author[1*]{Joonhyung Bae}

\affilblock[1]{Graduate School of Culture Technology, KAIST (Korea Advanced Institute of Science and Technology)}
              {Daejeon, South Korea}
              {[e-mail: jh.bae@kaist.ac.kr] \quad [ORCID: \href{https://orcid.org/0000-0001-5933-4302}{0000-0001-5933-4302}]}
\affilblock[*]{Corresponding author: Joonhyung Bae}{}{}
\date{}

\maketitle

\TIISarticledates{Received XXXX; revised XXXX; accepted XXXX; \\ published XXXX}

\begin{abstract}
The global landscape of art-technology institutions -- festivals, biennials, research labs, conferences, and hybrid organizations -- has grown increasingly diverse, yet systematic frameworks for analyzing their multidimensional characteristics remain scarce.
This paper proposes ASTRA (Art-technology Institution Spatial Taxonomy and Relational Analysis), a computational methodology that combines an eight-axis conceptual framework (Curatorial Philosophy, Territorial Relation, Knowledge Production Mode, Institutional Genealogy, Temporal Orientation, Ecosystem Function, Audience Relation, and Disciplinary Positioning) with a text-embedding and clustering pipeline to map 78 cultural-technology institutions into a unified analytical space.
Each institution is characterized through qualitative descriptions along the eight axes, which are encoded via E5-large-v2 sentence embeddings and quantized through a word-level codebook into TF-IDF feature vectors.
The reduction in dimensionality using UMAP, followed by agglomerative clustering (Average linkage, $k{=}10$), yields a composite evaluation score of 0.825, a silhouette coefficient of 0.803 and a Calinski–Harabasz index of 11,196.
Non-negative matrix factorization (NMF) extracts ten latent topics, and a neighbor-cluster entropy measure identifies boundary institutions that bridge multiple thematic communities.
An interactive web-based visualization tool built with React enables stakeholders—curators, researchers, and policymakers—to explore institutional similarities, thematic profiles, and cross-disciplinary connections.
The results reveal coherent groupings such as an ``art-science hub'' cluster anchored by ZKM and ArtScience Museum, an ``innovation \& industry'' cluster including Ars Electronica, transmediale, and S\'{o}nar, an ``ACM academic community'' cluster comprising TEI, DIS, and NIME, and an ``electronic music and media'' cluster including CTM Festival, MUTEK, and Sonic Acts.
This work contributes to a replicable and data-driven approach to institutional ecology in the cultural-technology sector and provides an open exploratory tool for the community.\footnote{Source code and data are available at \url{https://github.com/joonhyungbae/astra}.}
\end{abstract}

\keywords{art-technology institutions, interactive visualization, text embeddings, topic modeling, unsupervised clustering}

%% ━━━━━━━━━━━━━━━━━━━━━━━━━━━━━━━━━━━━━━━━━━━━━━━━━━━━━━━━━━━
%%  1  INTRODUCTION
%% ━━━━━━━━━━━━━━━━━━━━━━━━━━━━━━━━━━━━━━━━━━━━━━━━━━━━━━━━━━━
\section{Introduction}
\label{sec:intro}

The institutions that produce, curate, and disseminate media art and creative technology—festivals such as Ars Electronica (est. 1979), conferences such as ACM SIGGRAPH, and laboratories such as MIT Media Lab—constitute the institutional infrastructure of the global art-technology ecosystem.
Although computational methods have been successfully applied to cultural \emph{content}, games reviews~\cite{wang2020game_experience}, virtual-reality experiences, and interactive narratives, the institutional landscape that enables this content to emerge has received almost no computational attention.
Understanding how these institutions relate to each other, the thematic communities they form, and where cross-disciplinary bridges exist is essential for researchers, curators, and policy makers seeking to navigate this ecosystem.

Conventional approaches classify art-technology institutions by organizational form: festival, biennial, conference, museum, laboratory, residency, etc.
Although administratively convenient, such single-type labels capture only one facet of institutional identity.
Ars Electronica, for example, simultaneously functions as a festival, a museum (Ars Electronica Center), a research laboratory (Futurelab) and an international prize platform (Prix Ars Electronica).
Similarly, transmediale operates as a biennial, publishing platform, and residency program; ZKM integrates a museum, a research institute and a production house under one umbrella.
These multidimensional roles suggest that a concept-driven analytical framework is needed to go beyond categorical labels.

Advances in NLP and unsupervised learning now enable computational analysis of large-scale cultural data~\cite{manovich2020cultural,stoltz2021cultural_cartography}.
NLP-driven clustering and topic modeling have been applied to cultural \emph{content} -- game reviews~\cite{wang2020game_experience,sim2025cross_platform}, film corpora~\cite{chen2022animated_movie}, and player behavior~\cite{delalleau2012beyond_skill} -- yet never to \emph{institutions themselves}, leaving institutions computationally unexplored.

This paper introduces ASTRA, a computational framework for mapping this institutional landscape.
We propose an eight-axis conceptual framework and a text-embedding and clustering pipeline that transforms qualitative axis descriptions into a shared analytical space amenable to unsupervised clustering, topic modeling, and boundary analysis based on entropy.
Results are delivered through the APESuite Explorer, an interactive web application enabling non-technical stakeholders to explore institutional similarities and thematic communities.

The main contributions of this work are as follows.
\begin{enumerate}[nosep]
  \item \textbf{An eight-axis conceptual framework} for characterizing art-technology institutions beyond single-type labels, enabling multidimensional institutional analysis.
  \item \textbf{A text-embedding and codebook pipeline} that converts qualitative institutional descriptions into quantitative feature vectors through E5-large-v2 sentence embeddings and word-level codebook quantization—the first application of modern sentence-level embeddings to institutional profiles in the cultural-technology domain.
  \item \textbf{A clustering and topic-modeling methodology} that identifies 10 coherent institutional groupings and 10 latent themes, revealing cross-disciplinary connections invisible to conventional typologies.
  \item \textbf{An interactive web-based application} (APESuite Explorer) built on the APESuite platform makes these analytical results accessible to curators, researchers, and policy makers, supporting institutional benchmarking, ecosystem mapping, and cross-disciplinary discovery; a formal user evaluation is left for future work.
\end{enumerate}

%% ━━━━━━━━━━━━━━━━━━━━━━━━━━━━━━━━━━━━━━━━━━━━━━━━━━━━━━━━━━━
%%  2  RELATED WORK
%% ━━━━━━━━━━━━━━━━━━━━━━━━━━━━━━━━━━━━━━━━━━━━━━━━━━━━━━━━━━━
\section{Related Work}
\label{sec:related}

This section reviews computational cultural analysis, text embeddings, clustering, interactive visualization, and research gaps addressed by this work.

\subsection{Computational Analysis of Cultural and Institutional Ecosystems}
\label{sec:rel_ecosystem}

The computational study of cultural institutions and creative ecosystems has evolved along two complementary tracks: macro-scale cultural analytics and domain-specific institutional mapping.

\paragraph{Cultural analytics.}
Manovich's Cultural Analytics program~\cite{manovich2020cultural} established the foundational paradigm of treating culture as data, advocating for visualization and pattern recognition as tools for analyzing large cultural datasets.
This program pioneered computational analysis of millions of cultural artifacts but focused on \emph{content} rather than the institutions producing and circulating it.
Stoltz and Taylor~\cite{stoltz2021cultural_cartography} extended this vision by proposing two approaches for constructing ``cultural cartographies'' in word-embedding space, demonstrated on immigration discourse in U.S.\ media.
However, it maps \emph{media corpora} rather than institutional profiles.

\paragraph{Domain-specific institutional mapping.}
Several projects have cataloged institutions in specific cultural-technology sub-domains.
The BioArtlas project~\cite{bae2025bioartlas} systematically labeled and cataloged bioart works and their creators, providing structured metadata for computational analysis.
Separately, Yetisen et al.~\cite{yetisen2015bioart} provided a broad review of the field of bioart, reviewing its materials, techniques, and cultural implications.

\subsection{Theoretical Foundations for Institutional Characterization}
\label{sec:rel_theory}

Four bodies of scholarship inform the conceptual axes adopted in this paper (\Cref{tab:axis_theory} summarizes the mapping from theory to axis).
Bourdieu's theory of cultural production~\cite{bourdieu1993field,bourdieu2018forms} argues that institutional identity is constituted by the \emph{composition} of capital---economic, cultural, and symbolic---rather than by any single attribute, motivating a multi-axis framework that captures position-taking within the field.
DiMaggio and Powell~\cite{dimaggio1983iron} identify the coercive, mimetic, and normative pressures that drive organizations toward structural similarity (isomorphism); a computational taxonomy can detect where institutions \emph{resist} isomorphism, requiring axes sensitive to institutional genealogy and ecosystem function.
Quinn~\cite{quinn2005arts} and N\'{e}grier~\cite{negrier2015festivalisation} distinguish ``festival of a place'' (linked to local identity) from ``festival in a place'' (spatially mobile brands), informing the territorial relationship and audience relationship axes.
Hoetzlein~\cite{hoetzlein2023knowledge} proposes that new-media-art communities organize around ``knowledge cultures'' defined by shared values rather than media categories, informing the Knowledge Production Mode axis; complementarily, Grau~\cite{grau2007media_art} identifies media art's vulnerability to technological obsolescence, motivating the Temporal Orientation axis.
These perspectives converge on the idea that art-technology institutions must be characterized along multiple, theoretically motivated dimensions; the eight-axis framework (\Cref{sec:data}) operationalizes this idea.

\subsection{Text Embeddings and Topic Modeling for Cultural Data}
\label{sec:rel_nlp}

\paragraph{Sentence-level embeddings.}
Pre-trained sentence embeddings have transformed textual representation for computational cultural analysis.
Reimers and Gurevych~\cite{reimers2019sentence} introduced Sentence-BERT, enabling efficient semantic similarity computation via Siamese BERT networks.
More recent models -- including the BGE family~\cite{xiao2023bge,chen2024bgem3} and the GTE family~\cite{zhang2024mgte} -- achieve state-of-the-art performance on the Massive Text Embedding Benchmark (MTEB) and are particularly suited for domain-specific retrieval and clustering tasks.
Dual-model approaches can improve robustness by capturing complementary semantic aspects.

\paragraph{Topic modeling.}
Topic modeling is a well-established tool for extracting latent themes from textual corpora.
In the cultural analytics literature, LDA-based topic modeling has been applied to game and film reviews~\cite{wang2020game_experience,sim2025cross_platform,chen2022animated_movie}.
Non-negative matrix factorization (NMF)~\cite{lee1999nmf} offers an alternative that produces sparser, more interpretable topic--term distributions and is particularly suited to short-text and TF-IDF-weighted input, motivating its adoption in the present study.

\paragraph{Codebook quantization.}
The concept of discretizing continuous representations via learned codebooks originates in Van den Oord et al.'s work on Vector Quantized Variational Autoencoders (VQ-VAE)~\cite{van2017vqvae}.

The present paper applies codebook quantization to \emph{text embeddings} of institutional descriptions, a methodological combination that has not been demonstrated previously.

\subsection{Unsupervised Clustering and Interactive Visualization for Cultural Data}
\label{sec:rel_cluster_viz}

\paragraph{Clustering and dimensionality reduction.}
Unsupervised clustering is fundamental for discovering latent structure in high-dimensional cultural data.
UMAP~\cite{mcinnes2018umap} has emerged as the preferred dimensionality-reduction technique for cultural-data visualization, preserving both local and global structure more effectively than t-SNE~\cite{vandermaaten2008tsne} at comparable computational cost.

In related domains, clustering methods have been applied to player segmentation~\cite{delalleau2012beyond_skill} and emotion classification in movie imagery.

\paragraph{Interactive visualization for cultural data.}
Windhager et al.~\cite{windhager2019visualization_survey} provide the definitive survey of visualization systems for digital cultural heritage, reviewing over 75 tools categorized by visual granularity, interaction methods, and data types.
Nearly all reviewed systems focus on \emph{artifact-level} visualization: paintings, coins, manuscripts, and music scores.

Abramson and Nian~\cite{abramson2025cultural} introduced the Cultural Mapping and Pattern Analysis Toolkit (CMAP), integrating text analysis, clustering, dimensionality reduction, and visualization for cultural data, sharing the goal of making computational cultural analysis accessible.

\subsection{Research Gaps and Positioning}
\label{sec:rel_gaps}

The preceding review reveals three cross-cutting gaps that the present paper uniquely addresses:

\begin{enumerate}[nosep]
  \item \textbf{Institutions as units of analysis.}
    Across all surveyed areas, computational cultural analytics predominantly treats individual artifacts (artworks, games, music, films) or academic publications as units of analysis.
 To our knowledge, no previous study uses text embeddings based on NLP to computationally map art-technology \emph{institutions} as the primary analytical unit, despite the rich theoretical apparatus available for organizational analysis (\Cref{sec:rel_theory}).

  \item \textbf{Modern sentence embeddings with codebook quantization for institutional mapping.}
    Although transformer-based NLP has been applied to cultural-heritage texts and cultural cartography has been theorized with word embeddings~\cite{stoltz2021cultural_cartography}, no study uses state-of-the-art sentence embeddings (E5, BGE, GTE) combined with codebook quantization to map art-technology organizations in semantic space.

  \item \textbf{An integrated multi-method pipeline for information systems.}
    Existing work applies individual methods—bibliometrics, topic modeling, geographic clustering, or visualization~\cite{windhager2019visualization_survey}—in isolation.
    No prior study integrates text embedding, codebook quantization, clustering, topic modeling, and interactive visualization into a single end-to-end framework; this gap is especially pronounced in information-systems research, where knowledge management and web-based cultural heritage have been addressed, but institutional mapping driven by NLP with an interactive exploration tool remains absent.
\end{enumerate}

\Cref{tab:related_summary} provides a comparative summary of related studies and the positioning of this work.

\begin{table*}[t]
  \centering
  \caption{Comparative summary of closely related studies and the present work. Columns indicate the unit of analysis, the text-representation method, the clustering approach, topic modeling, interactive visualization, and the target domain.}
  \label{tab:related_summary}
  \scriptsize
  \renewcommand{\arraystretch}{1.15}
  \setlength{\tabcolsep}{0pt}
  \begin{tabularx}{\textwidth}{@{}
    >{\raggedright\arraybackslash}p{2.9cm}@{\hspace{8pt}}
    >{\centering\arraybackslash}p{1.3cm}@{\hspace{8pt}}
    >{\raggedright\arraybackslash}p{3.1cm}@{\hspace{8pt}}
    >{\centering\arraybackslash}p{1.35cm}@{\hspace{8pt}}
    >{\centering\arraybackslash}p{1.15cm}@{\hspace{8pt}}
    >{\centering\arraybackslash}p{1.45cm}@{\hspace{8pt}}
    >{\raggedright\arraybackslash}X
  @{}}
    \toprule
    \textbf{Study} & \textbf{Unit} & \textbf{Representation} & \textbf{Clustering} & \textbf{Topics} & \textbf{Viz Tool} & \textbf{Domain} \\
    \midrule
    Manovich~\cite{manovich2020cultural} & Artifacts & Image/text features & --- & --- & Custom & General culture \\
    Stoltz \& Taylor~\cite{stoltz2021cultural_cartography} & Media corpora & Word2vec & --- & --- & --- & Immigration disc. \\
    Wang \& Goh~\cite{wang2020game_experience} & Game reviews & BoW / TF-IDF & --- & LDA & --- & Entertainment \\
    Sim et~al.~\cite{sim2025cross_platform} & Game reviews & BoW / TF-IDF & --- & LDA & --- & Entertainment \\
    Yetisen et~al.~\cite{yetisen2015bioart} & Survey & Metadata catalog & --- & --- & Web catalog & Bioart \\
    Bae~\cite{bae2025bioartlas} & Artworks & Dual LLM emb. + codebook & Agglom. & --- & React & Bioart \\
    Abramson \& Nian~\cite{abramson2025cultural} & Cultural data & Text analysis & $k$-means & LDA & Toolkit & General culture \\
    \midrule
    \textbf{Present work} & \textbf{Institutions} & \textbf{E5-large-v2} & \textbf{Agglom.} & \textbf{NMF} & \textbf{React} & \textbf{Art-technology} \\
    \bottomrule
  \end{tabularx}
\end{table*}

%% ━━━━━━━━━━━━━━━━━━━━━━━━━━━━━━━━━━━━━━━━━━━━━━━━━━━━━━━━━━━
%%  3  METHODOLOGY
%% ━━━━━━━━━━━━━━━━━━━━━━━━━━━━━━━━━━━━━━━━━━━━━━━━━━━━━━━━━━━
\section{Methodology}
\label{sec:method}

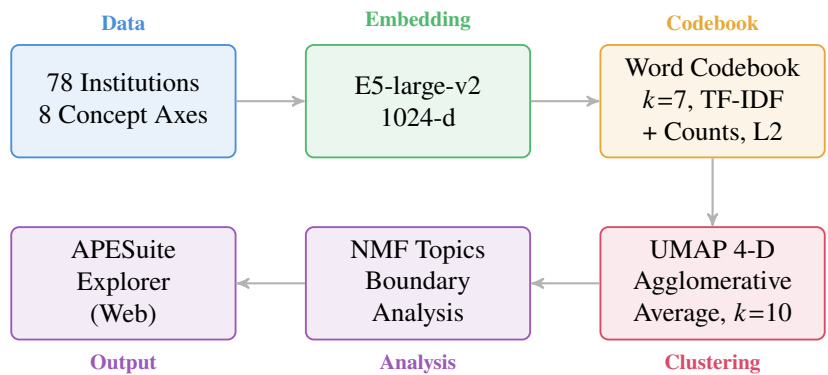
\begin{figure}[t]
  \centering
  \definecolor{stgA}{HTML}{4A90D9}
  \definecolor{stgB}{HTML}{50B86E}
  \definecolor{stgC}{HTML}{E8A838}
  \definecolor{stgD}{HTML}{D94A68}
  \definecolor{stgE}{HTML}{9B59B6}
  \begin{tikzpicture}[
    node distance=0.45cm and 0.9cm,
    every node/.style={font=\small},
    block/.style 2 args={
      rectangle, rounded corners=3pt,
      draw=#1, fill=#1!12,
      line width=0.7pt,
      inner xsep=5pt, inner ysep=2pt,
      execute at begin node=\begin{minipage}[c][1.35cm][c]{#2}\centering,
      execute at end node=\end{minipage}
    },
    arr/.style={-stealth', thick, draw=gray!60},
    slab/.style={font=\scriptsize\bfseries, text=#1},
  ]
  \node[block={stgA}{2.6cm}] (data) {78 Institutions\\8 Concept Axes};
  \node[slab=stgA, above=0.05cm of data] {Data};

  \node[block={stgB}{2.6cm}, right=0.9cm of data] (emb) {E5-large-v2\\$1024$-d};
  \node[slab=stgB, above=0.05cm of emb] {Embedding};

  \node[block={stgC}{2.6cm}, right=0.9cm of emb] (cb) {Word Codebook\\$k{=}7$, TF-IDF\\+ Counts, L2};
  \node[slab=stgC, above=0.05cm of cb] {Codebook};

  \node[block={stgD}{2.6cm}, below=0.9cm of cb] (clust) {UMAP 4-D\\Agglomerative\\Average, $k{=}10$};
  \node[slab=stgD, below=0.05cm of clust] {Clustering};

  \node[block={stgE}{2.6cm}, left=0.9cm of clust] (ana) {NMF Topics\\Boundary\\Analysis};
  \node[slab=stgE, below=0.05cm of ana] {Analysis};

  \node[block={stgE}{2.6cm}, left=0.9cm of ana] (web) {APESuite\\Explorer\\(Web)};
  \node[slab=stgE, below=0.05cm of web] {Output};

  \draw[arr] (data) -- (emb);
  \draw[arr] (emb) -- (cb);
  \draw[arr] (cb) -- (clust);
  \draw[arr] (clust) -- (ana);
  \draw[arr] (ana) -- (web);
  \end{tikzpicture}
  \caption{Overview of the ASTRA processing pipeline.
    Qualitative axis descriptions for 78 institutions are encoded through E5-large-v2 sentence embeddings, quantized via a word-level codebook, and clustered using UMAP and agglomerative clustering.
    NMF topic modeling and entropy-based boundary analysis are applied post-clustering, and results are served through an interactive web visualization.}
  \label{fig:pipeline}
\end{figure}

\Cref{fig:pipeline} provides an overview of the ASTRA pipeline, which proceeds through five stages: (1) data collection and annotation of the conceptual axis, (2) text embedding with E5-large-v2, (3) construction of the word-level codebook and quantization of features, (4) reduction and clustering of dimensions with algorithm comparison, and (5) modeling and boundary analysis of topics.

\subsection{Data Collection and Conceptual Framework}
\label{sec:data}

\subsubsection{Institution Selection}
\label{sec:data_selection}

The data set comprises 78 institutions that span the global art-technology sector, including festivals, biennials, conferences, museums, research centers, laboratories, universities, residency programs, award platforms, and hybrid organizations.
Institutions were selected to represent the diversity of the field, drawing from established art-technology databases (e.g. BioArtlas~\cite{bae2025bioartlas}), professional community networks (e.g. ISEA, Leonardo) and expert curation.
The data set covers institutions founded between 1837 and 2024 in 25 countries, with a concentration in Europe (55\%) and North America (25\%).

Each institution is classified by a primary type and a secondary type, yielding 14 primary categories: Conference~(17), Festival~(12), Center~(12), University~(8), Lab~(7), Biennial~(6), Residency~(5), Education~(3), Award~(3), and Other~(5).

\subsubsection{Axis Design Rationale}
\label{sec:axis_design}

An initial set of 13 candidate axes was derived from the theoretical foundations reviewed in \Cref{sec:rel_theory} and filtered through three criteria: (1)~\emph{discriminative power} -- axes with insufficient inter-institutional variance were removed; (2)~\emph{operational} adaptability -- each axis must be expressible as a short text amenable to sentence embeddings; and (3)~\emph{mutual independence} --overlapping axes were merged (e.g., ``Institutional Durability'' into Temporal Orientation, ``Funding Logic'' into Institutional Genealogy).
This yielded the final eight axes; the mapping from theory to axis is summarized in \Cref{tab:axis_theory}.

\subsubsection{Eight-Axis Framework}
\label{sec:eight_axes}

Each axis is described for every institution through a short English text (typically 15--40 words of comma-separated keywords and phrases) crafted by a domain expert with eight years of professional experience in the art-technology sector, based on publicly available institutional self-descriptions, mission statements, curatorial texts, and program archives.
For each axis, a pre-defined keyword pool of 9--12 terms was established from the theoretical literature to guide the annotation, with 2--4 terms selected per institution; this semi-structured approach follows a prior annotation methodology~\cite{bae2025bioartlas}.
Because the annotations draw from public source material and are constrained to a fixed keyword vocabulary, the annotator's role is closer to structured extraction than to subjective interpretation; moreover, the codebook quantization stage further absorbs surface-level phrasing differences by mapping semantically similar terms to the same codeword.

\begin{enumerate}[nosep]
  \item \textbf{Curatorial Philosophy}: artistic vision and intellectual position (techno-optimist vs.\ critically discursive).
  \item \textbf{Territorial Relationship}: relationship with geographic context, from ``festival of a place'' to ``festival in a place.''
  \item \textbf{Knowledge Production Mode}: how the institution generates and validates knowledge (artistic research, peer review, public education, commercial R\&D).
  \item \textbf{Institutional Genealogy}: historical origins --grassroots community, state policy, university, or commercial initiative.
  \item \textbf{Temporal Orientation}: emphasis on historical preservation, contemporary engagement, or speculative futures.
  \item \textbf{Ecosystem Function}: structural role (catalyst, incubator, platform, archive, marketplace).
  \item \textbf{Audience Relation}: engagement mode, from participatory co-creation to professional networking.
  \item \textbf{Disciplinary Positioning}: stance on disciplinary boundaries (art-science, humanities-engineering, antidisciplinary).
\end{enumerate}

\Cref{tab:axis_theory} provides a summary mapping of each axis to its theoretical foundation and the institutional dimensions it captures.

\begin{table}[t]
  \centering
  \caption{Mapping of the eight conceptual axes to their theoretical foundations.}
  \label{tab:axis_theory}
  \small
  \begin{tabularx}{\linewidth}{lXl}
    \toprule
    \textbf{Axis} & \textbf{Core Question} & \textbf{Theoretical Basis} \\
    \midrule
    Curatorial Philosophy      & What does the institution stand for?                  & Hoetzlein; Bourdieu \\
    Territorial Relation       & How does it relate to place?                          & Quinn; N\'{e}grier \\
    Knowledge Production       & What knowledge does it produce and how?                & Hoetzlein; UNESCO FCS \\
    Institutional Genealogy    & Where does it come from?                              & DiMaggio \& Powell \\
    Temporal Orientation       & How does it relate to time?                           & Grau \\
    Ecosystem Function         & What role does it play in the ecosystem?               & Hannan \& Freeman~\cite{hannan1977population} \\
    Audience Relation          & How does it engage audiences?                         & N\'{e}grier \\
    Disciplinary Positioning   & Where does it sit on the disciplinary spectrum?        & Shanken~\cite{shanken2002art_tech} \\
    \bottomrule
  \end{tabularx}
\end{table}

\subsection{Text Embedding}
\label{sec:embedding}

For each institution, the eight axis texts are individually encoded using E5-large-v2~\cite{wang2022e5} (1{,}024-dimensional), which achieves the highest clustering quality (silhouette 0.845 in codebook $k{=}5$) among the configurations evaluated in our embedding sweep (\Cref{tab:embedding_extended}).
The model employs mean-token pooling with L2 normalization.
The embeddings are concatenated across the eight axes to form the representation per institution.

We also evaluate alternative configurations, including BGE-M3~\cite{chen2024bgem3}, GTE-Qwen2~\cite{zhang2024mgte}, and their dual concatenation; the use of two architecturally complementary models serves two purposes:
(a) it captures a wider range of semantic features than any single model, and
(b) it provides robustness against model-specific biases, as suggested by prior work on combining complementary embedding models~\cite{reimers2019sentence}.
The extended embedding comparison (\Cref{tab:embedding_extended}) confirms that multiple configurations produce viable clustering outcomes, with E5-large-v2 achieving the highest quality and dual-model configurations capturing complementary semantic facets.
We also evaluate a lightweight Sentence-BERT baseline and traditional methods (Word2Vec~\cite{mikolov2013word2vec}, TF-IDF); the extended comparison is reported in \Cref{tab:embedding_extended}.

\subsection{Word-Level Codebook and Feature Quantization}
\label{sec:codebook}

Rather than using the raw high-dimensional embeddings directly for clustering, we construct a \emph{word-level codebook} that maps the continuous embedding space to a discrete, interpretable vocabulary of conceptual clusters.

\paragraph{Codebook construction.}
All axis texts are tokenized and each token is embedded using E5-large-v2.
PCA is applied to reduce the token embeddings to 42 dimensions (retaining 95.3\% variance), and agglomerative clustering with $k{=}7$ partitions the token space into seven word clusters (codewords).
The codebook size $k{=}7$ was selected via the embedding sweep: we evaluate downstream institution-level clustering quality across $k \in \{5,7,9,11,13\}$; E5-large-v2 attains a silhouette of 0.845 at $k{=}5$ and 0.825 at $k{=}7$.
We select $k{=}7$ over $k{=}5$ because at $k{=}5$ two codewords collapse semantically related but conceptually distinct term groups (e.g., ``innovation/industry'' and ``community/education'' merge into a single codeword), reducing the framework's ability to discriminate institutional profiles along these dimensions.
Each codeword represents a semantic concept group (e.g., one codeword may capture terms related to ``innovation,'' ``industry,'' and ``economic,'' while another captures ``critical,'' ``theoretical,'' and ``discourse'').

\paragraph{Feature quantization.}
Each institution is then represented as a feature vector by aggregating the codeword assignments of its tokens.
We compute both TF-IDF-weighted frequencies and raw count vectors across all eight axes, concatenate them, and apply L2 normalization.
This produces a 56-dimensional binary-sparse feature vector (8 axes $\times$ 7 codewords; density $\approx$ 2.7\%), which serves as the primary input for clustering.

The codebook approach offers three advantages:
(1) it reduces the feature dimensionality from 1{,}024 to 56 while preserving discriminative structure;
(2) it produces interpretable features, since each dimension corresponds to a known semantic cluster;
(3) it enables TF-IDF weighting, which suppresses ubiquitous terms and highlights distinctive vocabulary.
An ablation comparing the codebook pipeline with a raw-embedding pipeline (concatenated 8-axis E5-large-v2 $\to$ UMAP sweep $\to$ agglomerative) confirms the codebook's advantage: silhouette 0.630 vs.\ 0.344, Calinski--Harabasz 7{,}422 vs.\ 40, and Davies--Bouldin 0.553 vs.\ 0.895, justifying the quantization step.

\subsection{Dimensionality Reduction and Clustering}
\label{sec:clustering}

\paragraph{UMAP projection.}
The L2-normalized feature vectors are projected using UMAP~\cite{mcinnes2018umap} with a cosine distance.
For clustering, we used a 4-dimensional UMAP embedding with 10 nearest neighbors, minimum distance 0.0, and spread 1.0.
A separate 2-dimensional UMAP projection is used for visualization.

\paragraph{Algorithm comparison.}
We evaluate four clustering algorithms under a systematic hyperparameter sweep:
\begin{itemize}[nosep]
  \item \textbf{Agglomerative clustering} (Ward and Average linkage) with $k \in \{2,\ldots,20\}$.
  \item \textbf{$k$-means} (spherical, Lloyd algorithm) with $k \in \{2,\ldots,20\}$.
  \item \textbf{DBSCAN}~\cite{ester1996dbscan} with $\epsilon$ selected via the knee of the $k$-distance graph and $\mathrm{min\_samples} \in \{3,5,10,15\}$.
  \item \textbf{OPTICS}~\cite{ankerst1999optics} with $\xi \in \{0.01,0.05,0.1,0.15\}$ and $\mathrm{min\_samples} \in \{3,5,10,15\}$.
\end{itemize}

\paragraph{Composite evaluation metric.}
Each clustering result is evaluated in the post-UMAP space using a composite score that balances multiple quality criteria:
\begin{equation}
  S_{\text{composite}} = \alpha\,\hat{s}_{\text{sil}} + \beta\,\hat{s}_{\text{CH}} + \gamma\,\hat{s}_{\text{DB}} + \delta\,b_k - \lambda_1\,p_{\text{singleton}} - \lambda_2\,p_{\text{small}}
  \label{eq:composite}
\end{equation}
where $\hat{s}_{\text{sil}}$, $\hat{s}_{\text{CH}}$, and $\hat{s}_{\text{DB}}$ are the min-max normalized silhouette, Calinski--Harabasz and Davies--Bouldin scores, respectively.
Since lower DB values indicate better clustering, the DB score is inverted during normalization: $\hat{s}_{\text{DB}} = ({\text{DB}_{\max} - \text{DB}})/({\text{DB}_{\max} - \text{DB}_{\min}})$, so that higher $\hat{s}_{\text{DB}}$ values correspond to better results.
The granularity bonus $b_k$ rewards $k \in [k_{\min}, k_{\max}]$ via a linear ramp (slope $\eta{=}0.02$, $k_{\min}{=}5$, $k_{\max}{=}12$, max bonus~$0.14$); it equals zero for $k < k_{\min}$.
The penalty terms $p_{\text{singleton}} = |\{c : n_c {=} 1\}|/K_{\text{eff}}$ and $p_{\text{small}} = |\{c : n_c {<} 2\}|/K_{\text{eff}}$ discourage degenerate clusters, where $K_{\text{eff}}$ is the number of non-empty clusters.
The weights were fixed \emph{a priori} before any clustering experiments as $\alpha{=}0.30$, $\beta{=}0.25$, $\gamma{=}0.20$, $\delta{=}0.10$, $\lambda_1{=}0.10$, $\lambda_2{=}0.05$, reflecting the standard priority of cluster cohesion (silhouette) and separation (CH) in unsupervised evaluation; no post-hoc adjustments of these weights were performed.
A global sensitivity analysis (500 Dirichlet-sampled weight vectors, each constrained to $[0.05, 0.50]$) confirms that Agglomerative Average wins in 100\% of the tests ($k{=}8$ in 60.4\%, $k{=}10$ in 39.6\%).
The gap statistic~\cite{tibshirani2001gap} independently produces $k^{*}{=}8$ with a plateau over $k \in [8, 11]$, confirming that $k{=}10$ is statistically supported.
We selected $k{=}10$ for its finer-grained thematic resolution while retaining comparable metric quality.

\subsection{Topic Modeling and Boundary Analysis}
\label{sec:topics}

\paragraph{NMF topic extraction.}
Non-negative matrix factorization (NMF)~\cite{lee1999nmf} is applied to the codebook feature matrix.
The number of topics was selected by evaluating the reconstruction error, the diversity of the topic (the fraction of unique terms among the top-10 terms of all topics), and the mean inter-topic cosine similarity for $k \in \{3,\ldots,20\}$.
The reconstruction error curve exhibits a clear elbow around $k{=}8$--$10$, while the inter-topic similarity stabilizes below 0.02 for $k \geq 8$; we select $k{=}10$ as the configuration that balances low reconstruction error with high topic diversity.
Each resulting topic is characterized by its main contributing codewords and assigned a human-readable label (\Cref{tab:clusters}).

\paragraph{Boundary institution identification.}
Institutions that bridge multiple thematic communities are detected through \emph{neighbor-cluster entropy}.
For each institution, the neighbors with the highest $K_{\text{nn}}$ similarity are identified using cosine similarity in the L2-normalized codebook feature vectors.
The \emph{normalized} Shannon entropy of the cluster distribution among these neighbors quantifies how ``cross-disciplinary'' an institution is:
\begin{equation}
  H_i = \frac{-\sum_{c} p_{ic}\,\log\,p_{ic}}{\log\,|\mathcal{C}_i|}
  \label{eq:entropy}
\end{equation}
where $p_{ic}$ is the proportion of institution~$i$'s top-$K_{\text{nn}}$ neighbors that belong to cluster~$c$, and $|\mathcal{C}_i|$ is the number of distinct clusters represented among these neighbors.
Normalization by $\log|\mathcal{C}_i|$ maps $H_i$ to $[0, 1]$, where $H_i = 1$ indicates uniform distribution across clusters.
Institutions with $H_i$ close to 1 are listed as boundary institutions.

\subsection{Interactive Web Visualization}
\label{sec:viz}

To make the analytical results accessible to non-technical stakeholders, we built APESuite Explorer, an interactive web application hosted on the APESuite platform (\url{https://apesuite.org}).
The application is a single-page interface built with React 18, Vite, and Tailwind CSS.
The clustering pipeline exports a single JSON file containing UMAP coordinates, cluster labels, NMF topic weights, similarity scores, and institutional metadata; the React frontend consumes this file at build time and renders all visualizations client-side, requiring no backend server.
This design ensures lightweight deployment (static hosting) and reproducibility, since the visualization is fully determined by the pipeline output.
The tool provides the following interactive features:

\begin{itemize}[nosep]
  \item \textbf{2D scatter plot}: Institutions are rendered as SVG circles on UMAP coordinates, colored by cluster membership, with density contours delineating cluster boundaries. Users can pan, zoom (mouse wheel with zoom-to-pointer), and click to select.
  \item \textbf{Similarity links}: Selecting an institution highlights Bezier-curve links to its most similar neighbors, with line opacity proportional to similarity.
  \item \textbf{Filtering}: Users can filter by institution type, genre, and dominant NMF topic (theme), with a full-text search of names and descriptions.
  \item \textbf{Detail panel}: Displays institutional metadata, a thematic profile (top-3 topic weights), concept axis descriptions, and a list of similar institutions with similarity bars.
  \item \textbf{Multilingual support}: Full Korean and English localization via \texttt{react-i18next}.
\end{itemize}

%% ━━━━━━━━━━━━━━━━━━━━━━━━━━━━━━━━━━━━━━━━━━━━━━━━━━━━━━━━━━━
%%  4  RESULTS AND DISCUSSION
%% ━━━━━━━━━━━━━━━━━━━━━━━━━━━━━━━━━━━━━━━━━━━━━━━━━━━━━━━━━━━
\section{Results and Discussion}
\label{sec:results}

\subsection{Algorithm Comparison}
\label{sec:algo_compare}

\Cref{tab:algo} summarizes the best result achieved by each clustering algorithm.
Agglomerative clustering with Average linkage ($k{=}10$) achieves the highest composite score (0.825) and is selected as the primary configuration.
An interpretability comparison between the best Ward and Average configurations at comparable composite scores (Ward $k{=}13$ vs.\ Average $k{=}12$) shows that Average linkage yields higher mean intra-cluster topic focus (MITF 0.427 vs.\ 0.374) and lower cluster label entropy (1.535 vs.\ 1.639), further supporting Average linkage as the preferred algorithm independent of cluster count.
Although DBSCAN achieves the highest raw silhouette coefficient (0.998), it produces only two effective clusters, with 27.5\% of institutions classified as noise, making it impractical for interpretive analysis.
OPTICS provides a reasonable compromise (silhouette = 0.931, 10 clusters) but assigns 18.75\% of institutions as outliers.
The $k$-means is competitive (composite = 0.813, 13 clusters) but slightly inferior to agglomerative clustering in all normalized metrics.

\begin{table}[t]
  \centering
  \caption{Comparison of clustering algorithms (best configuration per algorithm).}
  \label{tab:algo}
  \begin{tabular}{lcccccc}
    \toprule
    \textbf{Algorithm} & \textbf{$k$} & \textbf{Composite} & \textbf{Silhouette} & \textbf{CH} & \textbf{DB} & \textbf{Noise\%} \\
    \midrule
    Agglomerative (Average) & 10 & \textbf{0.825} & 0.803 & 11{,}196 & 0.593 & 0.0\% \\
    OPTICS               & 10 & 0.818 & 0.931 & 18{,}934 & 0.348 & 18.75\% \\
    $k$-means            & 13 & 0.813 & 0.702 & 10{,}326 & 0.508 & 0.0\% \\
    DBSCAN               &  2 & ---   & 0.998 & ---       & ---   & 27.5\% \\
    \bottomrule
  \end{tabular}
  \par\smallskip
  {\footnotesize CH: Calinski--Harabasz index; DB: Davies--Bouldin index (lower is better).}
\end{table}

The stability of the agglomerative solution was further validated by bootstrap resampling, resulting in a mean adjusted Rand index (ARI) of 0.807 ($\sigma{=}0.117$) and a normalized mutual information (NMI) of 0.844 ($\sigma{=}0.044$), indicating robust cluster recovery across perturbations.

\begin{figure}[t]
  \centering
  \maybeincludegraphics[width=\linewidth]{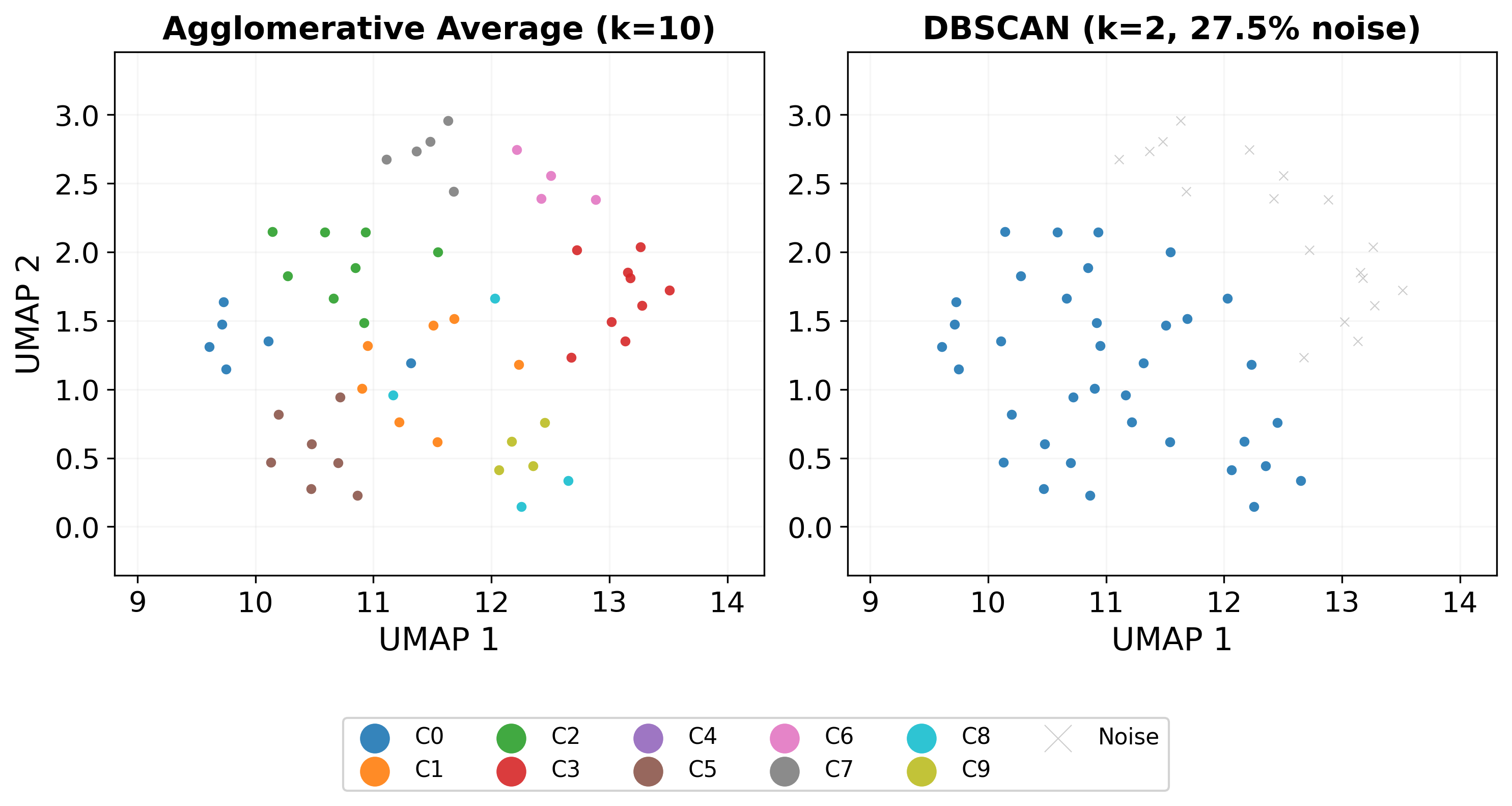}
  \caption{2D UMAP scatter plots comparing Agglomerative Average ($k{=}10$, left) and DBSCAN ($k{=}2$, right). Agglomerative clustering yields 10 interpretable groups, while DBSCAN produces only two effective clusters with 27.5\% of institutions classified as noise (gray).}
  \label{fig:algo_compare}
\end{figure}

\Cref{fig:algo_compare} contrasts the best (Agglomerative Average, $k{=}10$) and worst (DBSCAN, $k{=}2$, 27.5\% noise) clustering results on the same 2D UMAP projection.

\subsection{Ablation Studies}
\label{sec:ablation}

To validate key design decisions of the ASTRA pipeline, we performed three ablation experiments: (1)~an \emph{extended embedding comparison} that included E5-large-v2, GTE-Qwen2, dual-model configurations, a lightweight Sentence-BERT model, and traditional baselines (Word2Vec, TF-IDF); (2)~a \emph{leave-one-axis-out analysis} measuring each conceptual axis' contribution to clustering quality; and (3)~a \emph{negative control} using shuffled data to confirm that the observed clustering structure is not due to chance.

\paragraph{Extended embedding comparison.}
We evaluate embedding configurations under sweep-optimized conditions (codebook $k \in \{5,7,9,11,13\}$, UMAP 4D/8D, SVD, agglomerative/$k$-means, cluster $k \in \{8,\ldots,15\}$): E5-large-v2~\cite{wang2022e5}, GTE-Qwen2~\cite{zhang2024mgte}, GTE-Qwen2+SBERT dual, GTE-Qwen2+BGE dual, SBERT-Mini+BGE dual, Nomic~\cite{nussbaumer2024nomic}, BGE-M3~\cite{chen2024bgem3}, SBERT-Mini~\cite{reimers2019sentence}, and baselines (Word2Vec~\cite{mikolov2013word2vec}, TF-IDF).
\Cref{tab:embedding_extended} summarizes the results optimized for sweep.
E5-large-v2 achieves the highest silhouette coefficient (0.845 in the codebook $k{=}5$), Calinski--Harabasz (8{,}752) and the best Davies--Bouldin (0.494) among all configurations.
GTE-Qwen2+SBERT dual ranks second (silhouette 0.836 at $k{=}5$), followed by GTE-Qwen2+BGE dual (0.831 at $k{=}7$) and SBERT-Mini+BGE dual (0.828 at $k{=}7$).
TF-IDF standalone yields 0.139, and Word2Vec performs comparably to TF-IDF (silhouette below 0.2), confirming the substantial advantage of modern learned embeddings over traditional representations.
The codebook size $k{=}7$ is used in the main pipeline; E5-large-v2 with $k{=}5$ attains the sweep maximum (0.845), while $k{=}7$ with Average linkage yields the selected configuration (composite 0.825, silhouette 0.803) and is preferred for interpretability.

\begin{table}[t]
  \centering
  \caption{Embedding comparison (sweep-optimized: best codebook $k$, linkage, and cluster count per config; metrics computed in post-UMAP space). Parenthesized $k$ denotes codebook size. Values differ from \Cref{tab:algo} (which fixes cluster $k{=}10$, Average linkage) because each row reports its own sweep-optimal cluster count (ranging from $k{=}8$ to $k{=}13$).}
  \label{tab:embedding_extended}
  \begin{tabular}{lccc}
    \toprule
    \textbf{Configuration} & \textbf{Silhouette} & \textbf{CH Index} & \textbf{DB Index} \\
    \midrule
    E5-large-v2 (1{,}024-d, $k{=}5$)~\cite{wang2022e5} & \textbf{0.845} & \textbf{8{,}752} & \textbf{0.494} \\
    GTE-Qwen2+SBERT dual ($k{=}5$) & 0.836 & 10{,}171 & 0.503 \\
    GTE-Qwen2+BGE dual ($k{=}7$) & 0.831 & 13{,}962 & 0.556 \\
    SBERT-Mini+BGE dual ($k{=}7$) & 0.828 & 7{,}451 & 0.526 \\
    E5-large-v2 ($k{=}7$) & 0.825 & 14{,}317 & 0.556 \\
    Nomic ($k{=}9$) & 0.786 & 14{,}171 & 0.589 \\
    GTE-Qwen2 ($k{=}7$) & 0.778 & 11{,}177 & 0.564 \\
    SBERT-Mini ($k{=}9$)~\cite{reimers2019sentence} & 0.770 & 7{,}640 & 0.604 \\
    TF-IDF (standalone) & 0.139 & 4.3 & 1.68 \\
    Word2Vec~\cite{mikolov2013word2vec} & 0.118 & 3.8 & 1.72 \\
    \bottomrule
  \end{tabular}
\end{table}

\paragraph{Axis contribution analysis.}
A leave-one-axis-out study removes each of the eight conceptual axes in turn and re-runs the pipeline under identical conditions.
\Cref{tab:axis_contribution} reports silhouette and CH index computed in the codebook feature space (before UMAP); these values differ from the post-UMAP metrics in \Cref{tab:algo} because UMAP projection amplifies cluster separation.
Axes most critical for cluster cohesion ($\Delta_{\text{sil}}$) are \emph{Ecosystem Function} (+0.048) and \emph{Institutional Genealogy} (+0.047), while \emph{Knowledge Production Mode} ($\Delta_{\text{CH}} = -13.20$) contributes most to inter-cluster separation.
Three axes show negative $\Delta_{\text{sil}}$, indicating they introduce boundary complexity that enriches thematic resolution rather than indicating redundancy.
No single axis removal collapses the clustering, confirming that each dimension carries non-redundant information.

\begin{table}[t]
  \centering
  \caption{Leave-one-axis-out analysis. $\Delta_{\text{sil}}$ and $\Delta_{\text{CH}}$: changes relative to baseline (positive $\Delta_{\text{sil}}$ = removing the axis \emph{decreases} quality).}
  \label{tab:axis_contribution}
  \small
  \begin{tabular}{lcccc}
    \toprule
    \textbf{Removed Axis} & \textbf{Sil.} & $\Delta_{\text{sil}}$ & \textbf{CH} & $\Delta_{\text{CH}}$ \\
    \midrule
    (none --- baseline)           & 0.399 & ---     & 49.70 & --- \\
    \midrule
    Curatorial Philosophy         & 0.366 & +0.033  & 55.38 & $-5.67$ \\
    Territorial Relation          & 0.395 & +0.004  & 48.17 & +1.54 \\
    Knowledge Production Mode     & 0.418 & $-0.019$& 62.91 & $-13.20$ \\
    Institutional Genealogy       & 0.352 & +0.047  & 43.88 & +5.83 \\
    Temporal Orientation          & 0.416 & $-0.017$& 51.48 & $-1.78$ \\
    Ecosystem Function            & 0.352 & +0.048  & 56.37 & $-6.67$ \\
    Audience Relation             & 0.411 & $-0.012$& 51.44 & $-1.73$ \\
    Disciplinary Positioning      & 0.393 & +0.006  & 50.68 & $-0.98$ \\
    \bottomrule
  \end{tabular}
\end{table}

\begin{figure}[t]
  \centering
  \includegraphics[width=\linewidth]{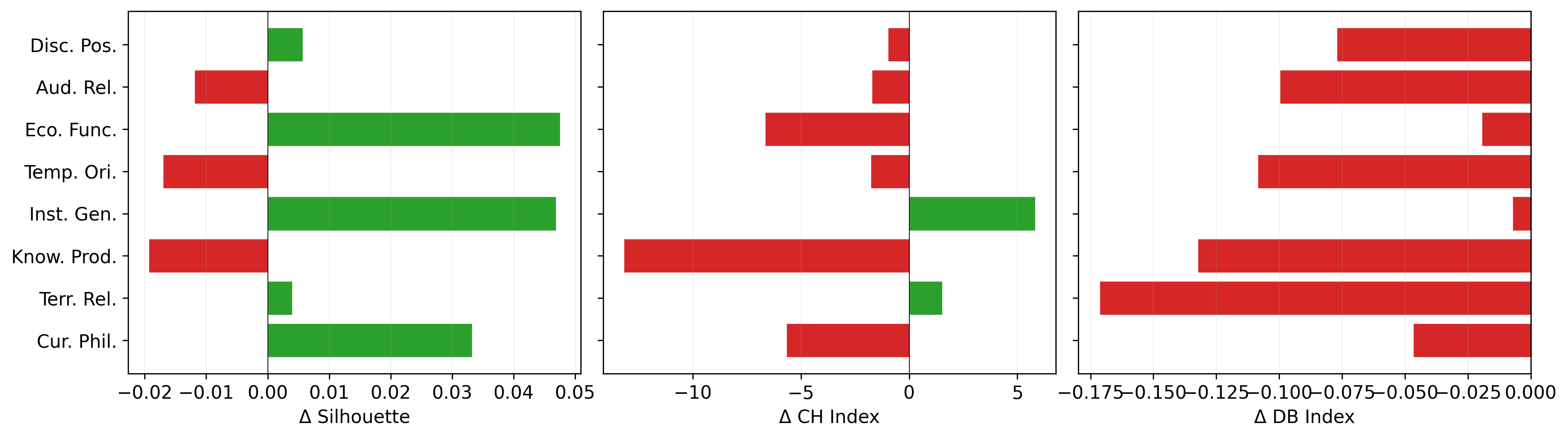}
  \caption{Axis contribution analysis (leave-one-axis-out). Each bar represents the change in the respective metric when the named axis is removed.}
  \label{fig:axis_ablation}
\end{figure}

\paragraph{Negative control.}
To rule out the possibility that the clustering structure arises from chance, we conducted negative-control experiments with shuffled data.
Three shuffle conditions are evaluated: (a) axis shuffle (permuting axis labels across institutions), (b) inter-institution shuffle (permuting institution IDs within each axis), and (c) token shuffle (permuting tokens within each axis description).
Evaluated in the original 56-dimensional codebook feature space (before UMAP projection), the pipeline achieves a silhouette score of 0.64; all three shuffle conditions yield significantly lower silhouette scores (axis shuffle: 0.59$\pm$0.03, $p < 10^{-30}$; inter-institution: 0.57$\pm$0.05, $p < 10^{-25}$; token shuffle: 0.56$\pm$0.10, $p < 10^{-11}$; two-sample $t$-test).
These results confirm that the observed cluster structure reflects a genuine semantic structure in institutional descriptions rather than spurious correlations.

\subsection{Cluster Analysis}
\label{sec:cluster_analysis}

\begin{figure}[t]
  \centering
  \maybeincludegraphics[width=1.0\linewidth]{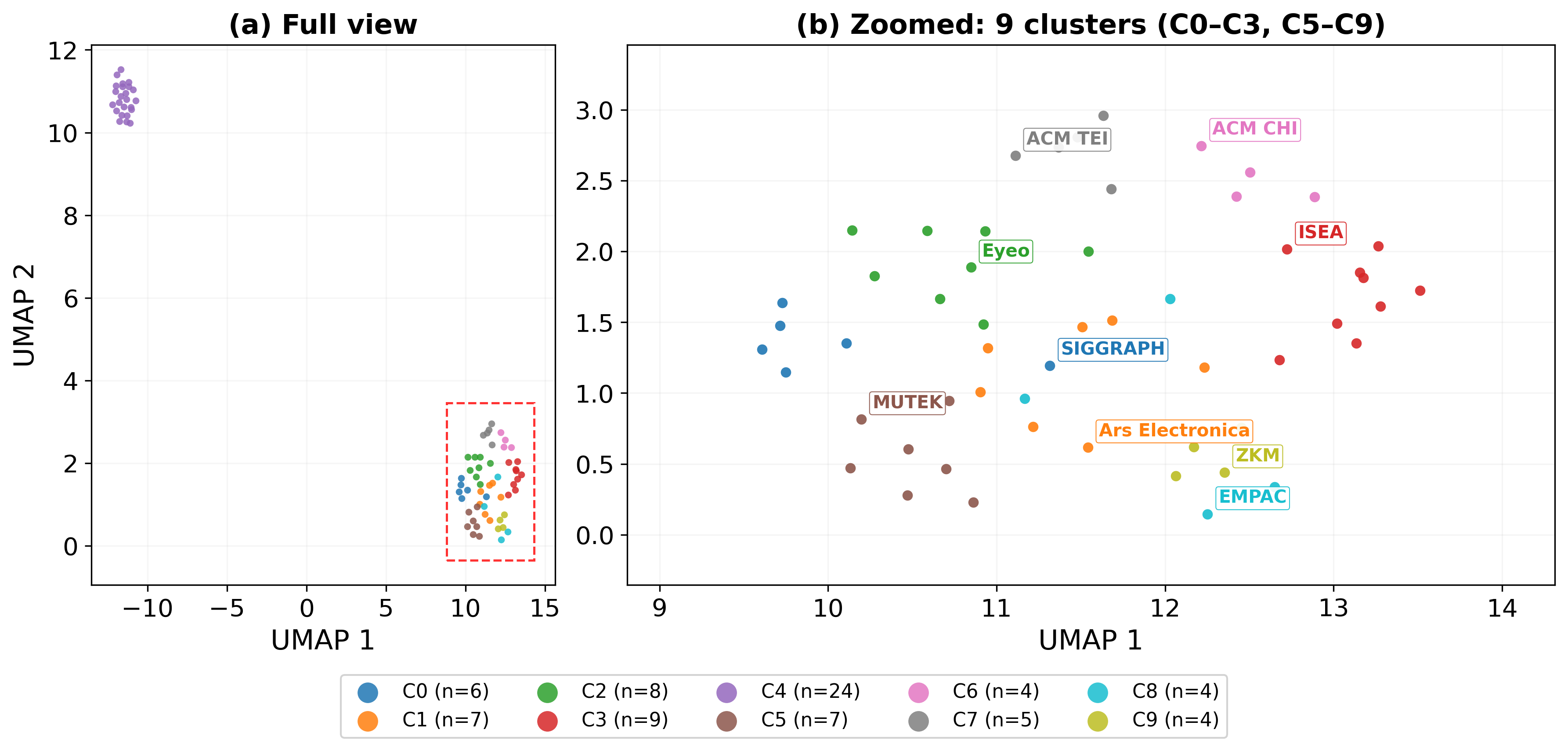}
  \caption{2D UMAP scatter plot of 78 institutions colored by cluster membership ($k{=}10$). (a)~Full view showing the spatial separation of Cluster~4; (b)~zoomed view of the nine remaining clusters with representative institution labels.}
  \label{fig:cluster_scatter}
\end{figure}

The 10 clusters identified by the agglomerative cluster reveal interpretable groupings of art-technology institutions.
\Cref{fig:cluster_scatter} shows the primary cluster map; institution types are distributed across cluster boundaries, confirming the value of concept-based analysis.
\Cref{tab:clusters} provides a summary of each group.

\begin{table}[t]
  \centering
  \caption{Summary of 10 clusters: size, founding year range, dominant topic, and representative institutions.}
  \label{tab:clusters}
  \small
  \begin{tabularx}{\linewidth}{ccclX}
    \toprule
    \textbf{Cl.} & \textbf{$n$} & \textbf{Years} & \textbf{Dominant Topic} & \textbf{Representative Institutions} \\
    \midrule
    0  &  6 & 1969--2014 & Computer Art \& Tech & BIAN, ACM SIGGRAPH, V2\_ Lab, e-flux, NEW INC \\
    1  &  7 & 1974--2015 & Innovation \& Industry & Ars Electronica, transmediale, S\'{o}nar, SXSW, ICMC \\
    2  &  8 & 1837--2012 & Creative Community & Eyeo, FILE, Resonate, EVA, ITP NYU \\
    3  &  9 & 1990--2013 & Biennials \& Confer. & NODE Forum, SeMA Mediacity Seoul, Biennale N\'{e}mo, ISEA \\
    4  & 24 & 1970--2017 & Mixed (Large) & Vivid Sydney, Piksel, SMC, Lumen Prize, Japan Media Arts \\
    5  &  7 & 1989--2011 & Music \& Sound Art & CTM, MUTEK, Elektra, Sonic Acts, ACM IMX \\
    6  &  4 & 1968--2011 & Digital Heritage & ACM CHI, Prix Ars Electronica, HeK, Leonardo \\
    7  &  5 & 1995--2007 & Participatory Design & ACM TEI, DIS, Multimedia, NIME, ISMIR \\
    8  &  4 & 1989--2013 & Art-Science \& Civic & Mapping, WRO, Chronus CAC, EMPAC \\
    9  &  4 & 1989--2016 & Art-Science Hubs & SIGGRAPH Asia, STARTS Prize, ZKM, ArtScience Museum \\
    \bottomrule
  \end{tabularx}
\end{table}

Several clusters merit detailed discussion:

\paragraph{Cluster~9: Art-science hubs.}
This cluster contains four major institutions anchored by the ZKM~\textbar ~ Center for Art and Media and the ArtScience Museum—organizations that integrate art and science within institutional frameworks.
The cluster's dominant topic is ``Art-Science \& Civic Integration''(topic~8, weight 0.273), reflecting these institutions' emphasis on bridging scientific research with public engagement.
SIGGRAPH Asia and the STARTS Prize (EU) round out the group.
We note that clusters with only four members (Clusters~6, 8, 9) should be interpreted with caution, as thematic characterizations derived from such small groups are inherently less stable than those from larger clusters.

\paragraph{Cluster~7: ACM academic community.}
The five institutions in this cluster -- ACM TEI, DIS, Multimedia (Art Program), NIME, and ISMIR -- represent the formal academic side of art-technology research.
Their dominant topic is ``Participatory Design Research'' (topic~2, weight 0.214) alongside ``Computer Art \& Technical Research'' (topic~5, weight 0.134), reflecting the peer-review, publication, and reproducibility ethos of the ACM community.

\paragraph{Cluster~5: Electronic music and media.}
This cluster groups seven institutions centered on sound, music, and media art: CTM Festival, MUTEK Montreal, Elektra Festival, Sonic Acts, and ACM IMX.
The dominant topic is ``Professional Digital Performance'' (topic~4, weight 0.244) alongside ``Music \& Sound Art'' (topic~7, weight 0.038), capturing the live-performance and immersive-experience orientation of these institutions.

\paragraph{Cluster~4: Large mixed cluster.}
The largest cluster (24 institutions) is also the most heterogeneous, encompassing institutions ranging from large-scale public events (Vivid Sydney, Japan Media Arts Festival) to research-oriented organizations (SMC, Lumen Prize).
Internal NMF analysis ($k{=}3$) reveals three coherent sub-groups: civic innovation hubs (Vivid Sydney, YCAM, teamLab; $n{=}9$), digital art community and support organizations (Piksel, Rhizome, SFPC; $n{=}9$), and scientific-experimental research institutions (IRCAM, Arts@CERN, SMC; $n{=}6$).

\subsection{Topic and Boundary Analysis}
\label{sec:topic_analysis}
\label{sec:boundary}

\begin{figure}[t]
  \centering
  \maybeincludegraphics[width=1.0\linewidth]{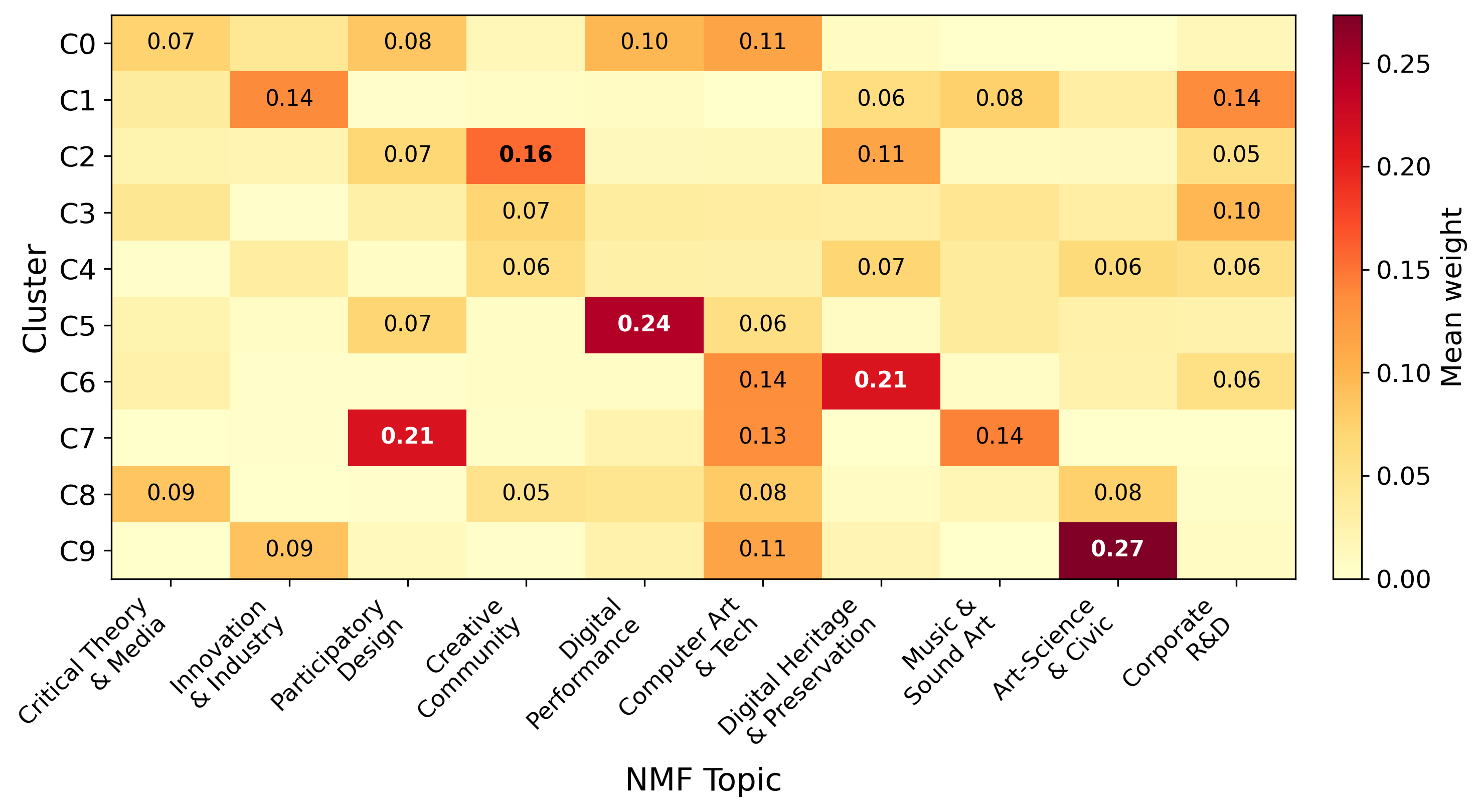}
  \caption{Cluster--topic heatmap. Each cell shows the mean NMF weight of the topic (column) within the cluster (row). Darker cells indicate stronger thematic associations.}
  \label{fig:topic_heatmap}
\end{figure}

\paragraph{Topic--cluster associations.}
\Cref{fig:topic_heatmap} presents the cluster--topic heat map, showing the mean weight of NMF topics for each group.
Several patterns emerge:

\begin{itemize}[nosep]
  \item \textbf{Topic~6 (Digital Heritage)} is heavily concentrated in Cluster~6 (weight 0.212), which contains historically oriented institutions such as ACM CHI, Prix Ars Electronica, HeK, and Leonardo.
  \item \textbf{Topic~2 (Participatory Design Research)} dominates Cluster~7 (weight 0.214), home to ACM conferences (TEI, DIS, NIME, ISMIR).
  \item \textbf{Topic~1 (Innovation \& Creative Industries)} peaks in Cluster~1 (weight 0.137), which includes festivals like Ars Electronica, transmediale, S\'{o}nar, and SXSW.
  \item \textbf{Topic~8 (ArtScience \& Civic Integration)} is strongest in Cluster~9 (weight 0.273), confirming its identity as the ArtScience hub (ZKM, ArtScience Museum).
  \item \textbf{Topic~4 (Professional Digital Performance)} is most pronounced in Cluster~5 (weight 0.244), reflecting the electronic music and media festival profile.
\end{itemize}

\paragraph{Boundary institutions.}

Entropy-based boundary analysis identifies 15 institutions with maximum or near-maximum normalized neighbor-cluster entropy at $k{=}10$ ($H_i \approx 1.0$), indicating that their top neighbors are distributed approximately uniformly across three or more distinct clusters.
A sensitivity analysis varying $K_{\text{nn}} \in \{3,5,7,10,15,20\}$ shows that the boundary set is moderately stable: at $K_{\text{nn}}{=}3$ the set inflates to 24 institutions (Jaccard~0.345 vs.\ $K_{\text{nn}}{=}10$), while $K_{\text{nn}} \geq 10$ converges (Jaccard~0.684--1.0), with 16 institutions appearing as boundaries across multiple $K_{\text{nn}}$ values.
\Cref{tab:boundary} lists selected boundary institutions.

\begin{table}[t]
  \centering
  \caption{Selected boundary institutions with high neighbor-cluster entropy.}
  \label{tab:boundary}
  \small
  \setlength{\tabcolsep}{3.5pt}
  \renewcommand{\arraystretch}{1.05}
  \begin{tabularx}{\linewidth}{@{}
    >{\raggedright\arraybackslash\hsize=1.55\hsize}X
    c
    >{\raggedright\arraybackslash}p{1.85cm}
    >{\raggedright\arraybackslash\hsize=0.75\hsize}X
    @{} }
    \toprule
    \textbf{Institution} & \textbf{Cl.} & \textbf{Type} & \textbf{Top-3 Neighbor Clusters} \\
    \midrule
    Ars Electronica Festival & 1 & Festival & 5, 1, 2 \\
    ZKM | Center for Art and Media & 9 & Center & 9, 4, 8 \\
    EMPAC & 8 & Residency & 8, 4, 9 \\
    Chronus Art Center (CAC) & 8 & Center & 8, 9 \\
    ACM TEI & 7 & Conference & 7, 3 \\
    SIGGRAPH Asia & 9 & Conference & 9, 4 \\
    STARTS Prize (EU) & 9 & Award & 9, 1 \\
    \bottomrule
  \end{tabularx}
\end{table}

In particular, Ars Electronica Festival -- despite being firmly situated in the ``innovation \& industry'' cluster (1) --has neighbors spanning the electronic music cluster (5), its own cluster (1) and the creative community cluster (2).
This quantitatively confirms its widely recognized role as a ``meta-institution'' that transcends conventional boundaries.
Similarly, ZKM bridges the art-science hub (9), the large mixed group (4), and the art-science \& civic cluster (8), reflecting its extensive programmatic scope as a museum, research center and production house.
More broadly, high neighbor-cluster entropy admits two interpretations: (a)~\emph{cross-pollinator} institutions that intentionally span multiple domains (e.g., Ars Electronica, ZKM), and (b)~\emph{identity-ambiguous} institutions whose profiles do not align strongly with any single thematic community.
Distinguishing between these cases requires examining whether the cross-cluster reach reflects deliberate programmatic breadth or a lack of consolidated identity -- a question the eight-axis profiles can inform.

\subsection{Interactive Visualization}
\label{sec:viz_results}

\begin{figure}[t]
  \centering
  \includegraphics[width=1.0\linewidth]{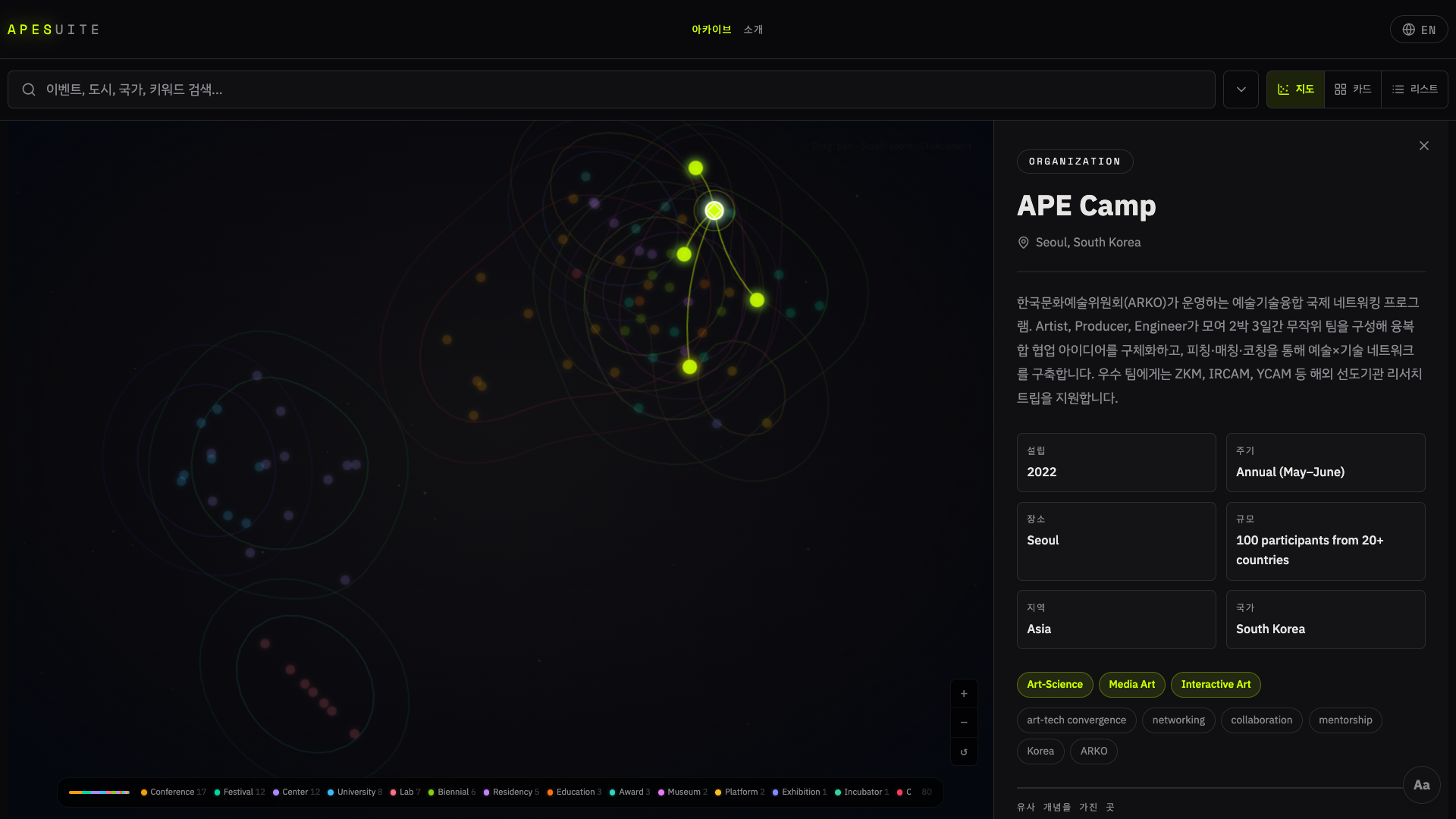}
  \caption{Screenshot of the APESuite Explorer web interface, showing the 2D scatter plot, selected institution detail panel with thematic profile, and similarity links.}
  \label{fig:web_screenshot}
\end{figure}

The APESuite Explorer provides a publicly accessible interface for exploring the clustering results.
\Cref{fig:web_screenshot} shows a representative screen capture.
Key usage scenarios include the following.

\begin{enumerate}[nosep]
  \item \textbf{Institutional comparison}: A curator seeking to understand how their festival relates to similar organizations can select it on the scatter plot and examine the similarity links and shared topics.
  \item \textbf{Ecosystem mapping}: A policymaker can filter by institution type (e.g., ``Lab'') and observe how research institutions are distributed across thematic clusters, informing funding strategy.
  \item \textbf{Cross-disciplinary discovery}: The boundary institution highlights reveal unexpected connections; for example, ACM IMX's proximity to both the electronic music cluster and the art-science hub suggests potential collaborative opportunities.
\end{enumerate}

\subsection{Discussion}
\label{sec:discussion}

The concept-based clustering reveals institutional groupings that cross conventional type-based taxonomies: transmediale (biennial) clusters with S\'{o}nar and SXSW (festivals) in the innovation \& industry cluster~(1), while CTM Festival and MUTEK form a distinct electronic music cluster~(5), reflecting shared curatorial philosophies rather than organizational format.
A codebook-level examination of the transmediale--SXSW pairing shows convergence on \emph{disciplinary positioning} (cosine similarity~1.0) despite opposing curatorial philosophies, capturing the structural insight that both function as professionalized platforms brokering between technology industries and cultural communities.
The distinction between Cluster~9 (art-science hubs) and Cluster~7 (ACM academic community) is similarly instructive: both engage with technology and art, but along fundamentally different knowledge production modes and audience relations.

\paragraph{Theoretical implications.}
Interpreted through the lenses of \Cref{sec:rel_theory}, three patterns emerge as hypotheses for future confirmatory research:
(1)~the separation of Cluster~1 from Cluster~3 -- both containing festivals and conferences -- is consistent with the Bourdieusian hypothesis that the capital composition structures the field~\cite{bourdieu1993field};
(2)~The uniformity of the cluster~7 is consistent with the normative isomorphism~\cite{dimaggio1983iron}, while the heterogeneity of the Cluster~4 may indicate unconsolidated isomorphic pressures;
(3)~Ars Electronica's bridging of three communities empirically exceeds the binary territorial model~\cite{quinn2005arts}.

\paragraph{Practical implications.}
The cluster map and thematic profiles offer actionable value for several stakeholder groups.
\emph{Curators and festival organizers} can use similarity links to identify peer institutions with complementary thematic strengths for co-programming; the spatial proximity of Cluster~5 (electronic music) and Cluster~9 (art-science hubs) mirrors real-world collaborations such as the joint programming between CTM Festival and transmediale in Berlin.
\emph{Funding agencies and policymakers} can overlay the cluster map with geographic data to identify under-supported institutional niches.
\emph{Researchers} entering the field can use boundary institutions (e.g. Ars Electronica, ZKM, EMPAC) as natural entry points for cross-disciplinary exploration.

\paragraph{Methodological considerations.}
Cluster boundaries are sensitive to UMAP hyperparameters; a systematic grid search confirms that $d_{\min}{=}0.0$--$0.1$ with $n_{\text{neighbors}}{=}10$ and 4D projection produce optimal results; full diagnostic plots are available in the project repository.
The large size of Cluster~4 (24~institutions) suggests that $k{=}10$ may slightly underfit the data; internal sub-topic analysis reveals three interpretable sub-groups (see \Cref{sec:cluster_analysis}).

%% ━━━━━━━━━━━━━━━━━━━━━━━━━━━━━━━━━━━━━━━━━━━━━━━━━━━━━━━━━━━
%%  5  CONCLUSION
%% ━━━━━━━━━━━━━━━━━━━━━━━━━━━━━━━━━━━━━━━━━━━━━━━━━━━━━━━━━━━
\section{Conclusion}
\label{sec:conclusion}

This paper presented ASTRA, a computational framework for mapping the global institutional landscape of art-technology.
By defining eight conceptual axes and employing an E5-large-v2 embedding pipeline with codebook quantization, we transformed qualitative institutional descriptions into a shared feature space amenable to unsupervised clustering and topic modeling.

The key findings are as follows.
\begin{enumerate}[nosep]
  \item The eight-axis framework and clustering pipeline reveal that art-technology institutions organize into 10 coherent thematic communities that cross-cut conventional type-based categories, for example, critically oriented biennials and commercial convergence festivals share structural roles despite oppositional intellectual orientations.
  \item NMF topic modeling identifies ten latent themes---from ``Art-Science \& Civic Integration'' to ``Music \& Sound Art''---that provide a thematic lens for understanding how the field is structured beyond organizational labels.
  \item Entropy-based boundary analysis identifies 15 institutions -- including Ars Electronica, ZKM, and EMPAC---that bridge multiple thematic communities, quantitatively confirming their widely recognized cross-disciplinary roles.
  \item The methodology achieves strong clustering quality (composite 0.825, silhouette 0.803), validated by gap statistic, bootstrap resampling, and global weight sensitivity analysis, and is delivered through the APESuite Explorer web interface for non-technical stakeholders.
\end{enumerate}

Whereas existing NLP-driven studies have mapped the landscape of cultural \emph{content} (games, films, social networks), ASTRA maps the \emph{institutional infrastructure} that enables this content to be created, curated, and disseminated.
By revealing the thematic structure and cross-disciplinary bridges within the art-technology ecosystem, the framework provides researchers, curators, and policy makers with a data-driven foundation for understanding how the field is organized at the institutional level.

\paragraph{Limitations.}
\begin{itemize}[nosep]
  \item \emph{Sample coverage}: 78 institutions across 25 countries provide meaningful diversity but are not exhaustive, with the Global South underrepresented. The Western-centric theoretical foundations (Bourdieu, DiMaggio \& Powell, Quinn, N\'{e}grier) may not generalize to ecosystems with different cultural and political logics.
  \item \emph{Single annotator}: Axis descriptions are based on English-language annotations by one domain expert with eight years of professional experience in the art-technology sector.
    Three design choices constrain annotator subjectivity: (1)~all descriptions are grounded in publicly available institutional self-descriptions, mission statements, and program archives rather than personal judgment; (2)~a pre-defined keyword pool of 9--12 terms per axis limits lexical variation; and (3)~the codebook quantization stage maps semantically similar terms to the same codeword, further absorbing surface-level phrasing differences.
    Nevertheless, inter-annotator agreement has not been formally measured, and the source material captures \emph{position-takings} rather than positions directly.
  \item \emph{Static snapshot}: The pipeline treats each institution as a fixed entity and does not capture programmatic shifts over time.
  \item \emph{No formal user evaluation}: The APESuite Explorer has been informally reviewed by practitioners but requires a task-based usability study.
\end{itemize}

\paragraph{Future work.}
Key directions include:
(a)~expanding the dataset to 200+ institutions with Global South representation;
(b)~introducing multi-annotator protocols with formal reliability metrics (Cohen's $\kappa$, Krippendorff's $\alpha$);
(c)~developing longitudinal analysis to track institutional trajectories;
(d)~conducting a formal user study of the APESuite Explorer;
and (e)~applying the domain-portable pipeline architecture to other institutional ecosystems (e.g., academic publishing, game-industry organizations), which requires only redesigning the conceptual axes and keyword pools while the remainder of the pipeline is domain-independent.

%% ━━━━━━━━━━━━━━━━━━━━━━━━━━━━━━━━━━━━━━━━━━━━━━━━━━━━━━━━━━━
%%  DECLARATION OF COMPETING INTEREST
%% ━━━━━━━━━━━━━━━━━━━━━━━━━━━━━━━━━━━━━━━━━━━━━━━━━━━━━━━━━━━
\section*{Declaration of competing interest}
The author declares that there are no known competing financial interests or personal relationships that could have appeared to influence the work reported in this document.

%% ━━━━━━━━━━━━━━━━━━━━━━━━━━━━━━━━━━━━━━━━━━━━━━━━━━━━━━━━━━━
%%  DATA AVAILABILITY
%% ━━━━━━━━━━━━━━━━━━━━━━━━━━━━━━━━━━━━━━━━━━━━━━━━━━━━━━━━━━━
\section*{Data availability}
The dataset of 78 institutional profiles, the clustering pipeline source code, and the APESuite Explorer web application are publicly
available at \url{https://github.com/joonhyungbae/astra}.

%% ━━━━━━━━━━━━━━━━━━━━━━━━━━━━━━━━━━━━━━━━━━━━━━━━━━━━━━━━━━━
%%  REFERENCES
%% ━━━━━━━━━━━━━━━━━━━━━━━━━━━━━━━━━━━━━━━━━━━━━━━━━━━━━━━━━━━

\bibliographystyle{TIIS}
\bibliography{reference}

\section*{Author Profile}
\noindent\mbox{}\par\vspace{-\baselineskip}

\begin{TIISbiography}[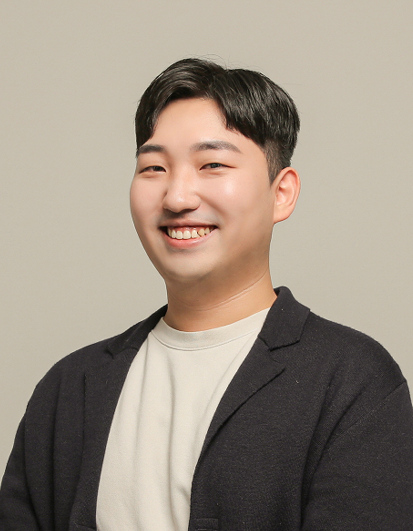]{Joonhyung Bae}
received the B.F.A. degree in Art \& Design from Korea University, Seoul, South Korea, in 2019, and the M.S. degree in Culture Technology from the Graduate School of Culture Technology, Korea Advanced Institute of Science and Technology (KAIST), Daejeon, South Korea, in 2021, where he is currently pursuing the Ph.D. degree under the supervision of Prof. Juhan Nam. He is a Research Assistant at the KAIST Music and Audio Computing Lab. His current research interests include multimodal artificial intelligence, human--computer interaction, interactive media systems, virtual reality, and real-time audio-visual information processing. He has published papers at top-tier international venues including NeurIPS, ACM SIGGRAPH Asia, ACM CHI, ISMIR, IEEE ISMAR, ISEA, and ICMC. He is a recipient of the CHI 2022 Student Game Competition Award in the Category of Transformative and Transgressive Play. He served as the Web Chair and Designer for ISMIR 2025 and as the Local Organization Chair for ISAIMP 2023. He has also served as a reviewer for ICASSP and Multimedia Tools and Applications.
\end{TIISbiography}

\end{document}